\documentclass[journal,transmag]{IEEEtran}

\usepackage{cite}
\usepackage{amsmath,amssymb,amsfonts}
\usepackage{algorithmic}
\usepackage{graphicx}
\usepackage{textcomp}
\usepackage{xcolor}
\usepackage{todonotes}
\usepackage{bm}
\usepackage{xspace}
\usepackage{booktabs}
\usepackage{tikz}
\usepackage{pgfplots}
\usepackage{inputenc}
\usepackage{standalone}
\usepackage{multirow}
\usepackage{pdfpages}
\usepackage[section]{placeins}
\usepackage{bm}
\usepackage{setspace}
\usepackage{soul}
\usepackage{cancel}
\usepackage{mathtools}
\usepackage[colorlinks, linkcolor={red!50!black},citecolor={blue!50!black},urlcolor={blue!80!black}]{hyperref}

\setstcolor{red}

\newcommand{\threeD}{\mbox{3-D}\xspace}
\newcommand{\twoD}{\mbox{2-D}\xspace}
\newcommand{\hs}{$\mathbf{H}_s$\xspace}
\newcommand{\hr}{$\mathbf{H}_r$\xspace}
\newcommand{\hrphi}{\hbox{H$_{\text{r}}$-$\phi$}\xspace}
\newcommand{\hra}{\hbox{H$_{\text{r}}$-A}\xspace}

\newlength\figH
\newlength\figW
\makeatletter
\edef\textFontName{\fontname\csname
	\f@encoding/\f@family/\f@series/\f@shape/\f@size\endcsname}

\makeatother

\begin{document}
	%
	\title{Efficient modeling of high temperature superconductors surrounded by magnetic components using a reduced \hbox{H-$\phi$} formulation}

	
\author{\IEEEauthorblockN{Alexandre Arsenault\IEEEauthorrefmark{1},
		Fr\'ed\'eric Sirois\IEEEauthorrefmark{1}, and
		Francesco Grilli\IEEEauthorrefmark{2}}
	\IEEEauthorblockA{\IEEEauthorrefmark{1}Polytechnique Montr\'eal, Montr\'eal, Canada}
	\IEEEauthorblockA{\IEEEauthorrefmark{2}Karlsruhe Institute of Technology, Karlsruhe, Germany}
	\thanks{Corresponding author: Alexandre Arsenault (alexandre-1.arsenault@polymtl.ca)}
}
	
	\IEEEtitleabstractindextext{%
		\begin{abstract}
			Although the H-formulation has proven to be one of the most versatile formulations used to accurately model superconductors in the finite element method, the use of vector dependent variables in non-conducting regions leads to unnecessarily long computation times. Additionally, in some applications of interest, the combination of multiple magnetic components interacting with superconducting bulks and/or tapes leads to large domains of simulation. In this work, we separate the magnetic field into a source and reaction field and use the \hbox{H-$\phi$} formulation to efficiently simulate a superconductor surrounded by magnetic bodies. We model a superconducting cube between a pair of Helmholtz coils and a permanent magnet levitating above a superconducting pellet. In both cases, we find excellent agreement with the H-formulation, while the computation times are reduced by factors of nearly three and four in \twoD and \threeD, respectively. Finally, we show that the H-$\phi$ formulation is more accurate and efficient than the H-A formulation in \twoD.
		\end{abstract}
		
		}

	\maketitle

	\IEEEdisplaynontitleabstractindextext

	%
	\IEEEpeerreviewmaketitle

	\section{Introduction}
	%
	%
	%
	%
	\IEEEPARstart{M}{odeling} the electromagnetic behavior of high temperature superconductor (HTS) devices has become an important practice in the development of novel applications. Many approaches have been proposed to simulate the electromagnetic performance of the highly nonlinear resistivity of superconductors (SC), but the most widely used is the finite element method (FEM). Even within the FEM, several different formulations are employed, such as the A-formulation (magnetic vector and electric scalar potential as the dependent variables) and the H-formulation (magnetic field as the dependent variable). In particular, the H-formulation has proven to be an exceptionally versatile formulation by successfully modeling numerous applications involving superconducting bulks and/or tapes\cite{grilli2013,zou2015,grilli2018,philippe2015,ainslie2014,stenvall2014,brambilla2007,sass2015,queval2018,zermeno2013a,shen2020,shen2020a}. In this formulation, the magnetic field is explicitly taken as the dependent variable, which makes the implementation simple and intuitive since no gauging or post-processing is required to obtain the magnetic field distribution. Nevertheless, the H-formulation unnecessarily adds degrees of freedom (DOFs) to the problem and requires an artificial resistivity in nonconducting domains, which degrades the matrix conditioning\cite{arsenault2021}.
	
	Recently, we showed that the magnetic scalar potential $\phi$ can be used to efficiently model the magnetic field in nonconducting regions surrounding magnetized bulk superconductors\cite{arsenault2021}. Using the magnetic scalar potential reduces the number of DOFs since the dependent variable is a scalar as opposed to the vector dependent variable used in the H-formulation. In addition, no artificial resistivity is needed when using $\phi$ in nonconducting domains. Therefore, the combination of H in superconducting domains and $\phi$ in nonconducting domains leads to decreased computation times, while preserving the full electromagnetic behavior of the SC. This formulation, called \hbox{H-$\phi$}, has only recently been used to model SCs\cite{arsenault2021,burger2019,dular2020}.
	
	In many applications of interest, multiple magnetic components interact together. For example, HTS bulks and tapes, permanent magnets (PM) and/or stranded coils are used in superconducting machines \cite{klaus2007,zhou2012} and in magnetic levitation systems \cite{bernstein2020}. In some cases, the interactions between the components are primarily unidirectional: the magnetic field produced by the PM or by the stranded coil interacts with the SC, while the field produced by the SC has negligible impact on the field source, either PM or coil. In such cases, it is oftentimes possible to separate the simulation into two parts: 1) the calculation of the source field \hs generated by the independent magnetic components, and 2) the computation of the reaction field \hr produced by the magnetizable bodies interacting with \hs. With this method, proposed in \cite{dular2004}, simulating the more computationally expensive HTS can be efficiently done by considering only a small region of air surrounding the HTS. Since we will be using the \hbox{H-$\phi$} formulation together with the method described above, we will refer to this formulation as a reduced \hbox{H-$\phi$} formulation, denoted \hrphi below.
	
	In this work, we describe how to implement the \hrphi formulation and use it to model two applications of interest. We first model the magnetization of a superconducting bulk between a pair of Helmholtz coils and validate our results with the benchmark~\#5 of the htsmodelling.com website\cite{htsmodeling}. We then simulate the magnetic levitation of a PM over a HTS bulk. In order to validate our results with the H-formulation, we simulate one of the levitation systems considered in \cite{grilli2018}.
	
	\section{Formulations}
	
	\subsection{H-$\phi$ formulation}
	
	The \hbox{H-$\phi$} formulation is a mixed formulation implemented by coupling the magnetic field \textbf{H} in conducting domains to the magnetic scalar potential $\phi$ in nonconducting domains. In conducting domains, the regular H-formulation, combining Ampere's and Faraday's laws, is used such that the governing equation is:
	\begin{equation}
	\quad \nabla\times \left(\rho\nabla\times\mathbf{H}\right) =-\mu_0\frac{\partial \mathbf{H}}{\partial t} \,,
	\label{eq:H}
	\end{equation}
	where $\rho$ is the resistivity and $\mu_0$ is the magnetic permeability of air. The nonlinear resistivity of the SC is modeled using the power law model \cite{rhyner1993}:
	\begin{equation}
	\rho=\frac{E_c}{J_c}\left(\frac{\|\mathbf{J}\|}{J_c}\right)^{n-1},
	\label{eq:rho}
	\end{equation}
	where $\mathbf{J}$ is the current density, $J_{\textnormal{c}}$ is the critical current density, $n$ is the power law exponent, and $E_c=1~\mu$V/cm.
	
	In nonconducting domains, Ampere's law states that \mbox{$\nabla\times\mathbf{H}=0$} when neglecting displacement currents, such that we can define the magnetic scalar potential as \mbox{$\mathbf{H}=-\nabla\phi$}. The equation generating the $\phi$ physics can be derived from the divergence-free condition of the magnetic flux density, \mbox{$\nabla\cdot\mathbf{B}=0$}. Since the magnetic flux density is related to the magnetic field through  \mbox{$\mathbf{B}=\mu_0\mathbf{H}$}, the governing equation in nonconducting domains is:
	\begin{equation}
	\nabla\cdot\nabla\phi=0.
	\label{eq:phi}
	\end{equation}

	The coupling between the two physics is explained in detail in \cite{arsenault2021}. Briefly, the tangential component of the fields are equated in the H physics, while the perpendicular components are equated in the $\phi$ physics. Through these two couplings, the full vector field is correctly defined at the boundary between the two physics.

	\subsection{Reduced H-$\phi$ formulation}
	
	As stated previously, we can model the electromagnetic behavior of many interacting magnetic objects by separating the simulation into the computation of source and reaction fields.
	
	The source field \hs can easily be obtained with any appropriate formulation. On the other hand, in order to expose the SC to the source field, we employ a modified H-formulation to calculate the reaction field \hr. The total magnetic field is given by $\mathbf{H}=\mathbf{H}_s+\mathbf{H}_r$, such that the standard H-formulation can be re-written in terms of source and reaction fields as:
	\begin{equation}
	\quad \nabla\times \left(\rho\nabla\times\mathbf{H}_r\right) =-\mu_0\frac{\partial}{\partial t}\Big(\mathbf{H}_r+\mathbf{H}_s\Big) \,,
	\label{eq:H_sep}
	\end{equation}
	where $\nabla\times\mathbf{H}_s=0$ inside the bulk volume, i.e. the source field is not associated with any currents inside the HTS domain. This modified H-formulation is only required in the superconducting domain, while the reaction field in all other domains is computed using \eqref{eq:phi}. We implement \eqref{eq:H_sep} with the help of the \textit{General Form PDE} module in COMSOL Multiphysics.

	\section{Application Examples}

	In this section, we explore two examples that can benefit from the \hrphi formulation. We first model the complete electromagnetic process of magnetizing a superconducting cube between a pair of Helmholtz coils. We then model the levitation of a PM above a HTS bulk by using the dynamic H-formulation model proposed in \cite{grilli2018}. All simulations are performed with COMSOL Multiphysics~5.5 on a workstation with an Intel(R) Xeon(R) E5-2690 processor @2.90~GHz  and 128~GB of random-access memory. 
	
	\subsection{Magnetization of HTS cube between a pair of Helmholtz coils}
	
\begin{figure}[t]
		\centering
		\includegraphics[width=\linewidth]{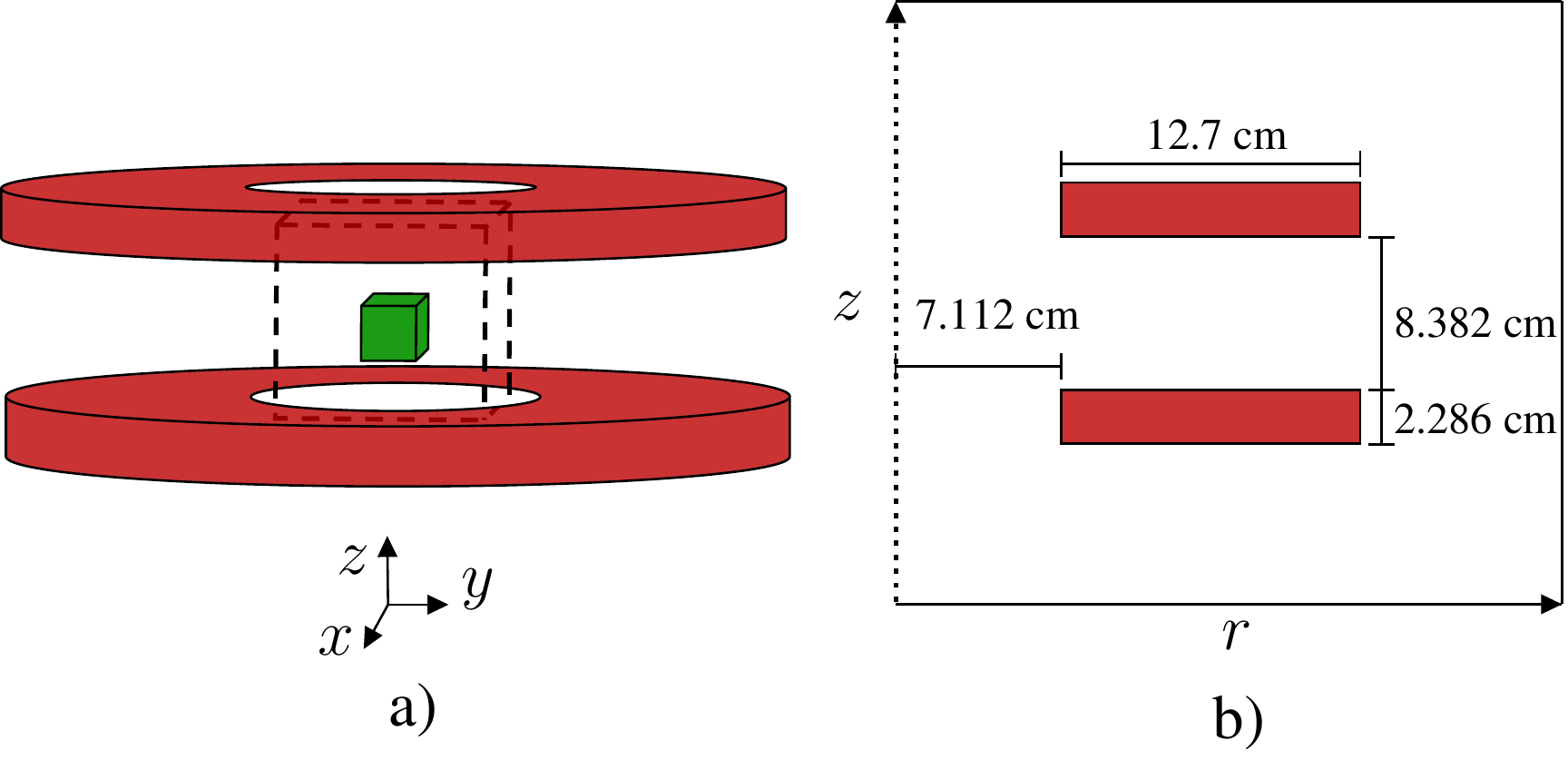}
		\caption{Simulated geometries considered in the magnetization portion of this work. a) \threeD geometry with cube-shaped superconductor between a pair of Helmholtz coils. The superconducting cube is drawn 3 times larger than its real size for readability purposes. See Fig.~\ref{fig:field} for its real dimensions. The dashed lines show the reduced domain simulated with the \hrphi formulation in this work and with the H-formulation in benchmark~\#5. b) \twoD axisymmetric geometry of the Helmholtz coils (cube not shown). The dotted line represents the symmetry axis. Illustrations created on \url{www.mathcha.io}.}
		\label{fig:geom}
	\end{figure}

	The geometry considered for the magnetization of a HTS cube placed between a pair of Helmholtz coils is shown in Fig.~\ref{fig:geom}a). A cube-shaped HTS of $a=1$~cm sides is magnetized between a pair of Helmholtz coils with inner radius of 7.112~cm, outer radius of 19.812~cm, thickness of 2.286~cm and inner separation of 8.382~cm. The dimensions of the coils are chosen to replicate our in-lab 5~T electromagnet, such that the field varies by only 0.01\% over the bulk volume. The nearly perfect uniformity of the field over the bulk volume enables us to compare our simulation results with benchmark~\#5 of the htsmodelling.com website\cite{htsmodeling}, where a uniform field is assumed. In this reference case, the uniform field is sinusoidal with an amplitude of 200~mT and a frequency of 50~Hz applied at the boundary of a 10~cm sides cube air domain surrounding the HTS. The critical current density and power law exponent are $1\times10^8$~A/m$^2$ and 100, respectively. 
	
	Although we previously found that higher order elements are more suitable for simulating the magnetization of bulk superconductors~\cite{arsenault2021}, we use linear elements for a fair comparison with the H-formulation of the benchmark. A total of 68,921 tetrahedral elements are used in the HTS domain, corresponding to the same amount used in the benchmark. The air domain surrounding the coils is 30~cm in radius and 60~cm in height.
	
	By separating the field produced by the coils (\hs) from the field produced by the bulk (\hr), we can exploit the \twoD axisymmetric nature of the coils\cite{dular2010}, as shown in Fig.~\ref{fig:geom}b). In addition, according to the Biot-Savart law, the magnetic field produced by the coils normalized by the current in the coils is a constant, such that
	\begin{equation}
		\frac{\mathbf{H}_1}{I_1}=\frac{\mathbf{H}_2}{I_2},
	\end{equation}
	where the subscripts represent an arbitrary value of magnetic field and current in the coils. Thus, by calculating a normalized, static source field, $\mathbf{H}_{s0}$, with a unitary current of 1~A$\cdot$turn in the coils, we can easily obtain the time-dependent source field
	\begin{equation}
		\mathbf{H}_s(t)=\mathbf{H}_{s0} N I(t),
		\label{eq:Hst}
	\end{equation}
	where $N$ is the number of turns in the coil and $I(t)$ is the time-dependent current imposed in a single turn of the coils to produce the magnetizing field.
	
	Consequently, $\mathbf{H}_s$ is calculated with a \twoD axisymmetric, static simulation of the coils with a normalized current of 1~A$\cdot$turn. The time-dependent nature of the magnetic field in the zero field cooled (ZFC) process is generated by \eqref{eq:Hst}, with $NI(t)$ being a sinusoidal function with an amplitude of $I_0=27.45$~kA$\cdot$turn in order to generate the field of 200~mT at the center of the pair of Helmholtz coils. Finally, the \twoD axisymmetric field is revolved to a \threeD field by using COMSOL's \textit{General extrusion} operator and through the coordinate transformation:
	\begin{align}
		H_{x}&=H_{r}\cos(\theta),\nonumber\\
		H_{y}&=H_{r}\sin(\theta),\\
		H_{z}&=H_{z},\nonumber
	\end{align}
	where $\theta$ is the azimuthal angle and H$_i$ corresponds to the magnetic field in the $i$ coordinate. This \threeD field is used as the source field \hs in \eqref{eq:H_sep}.
	
\begin{figure}[t]
		\centering
		\includegraphics[width=\linewidth]{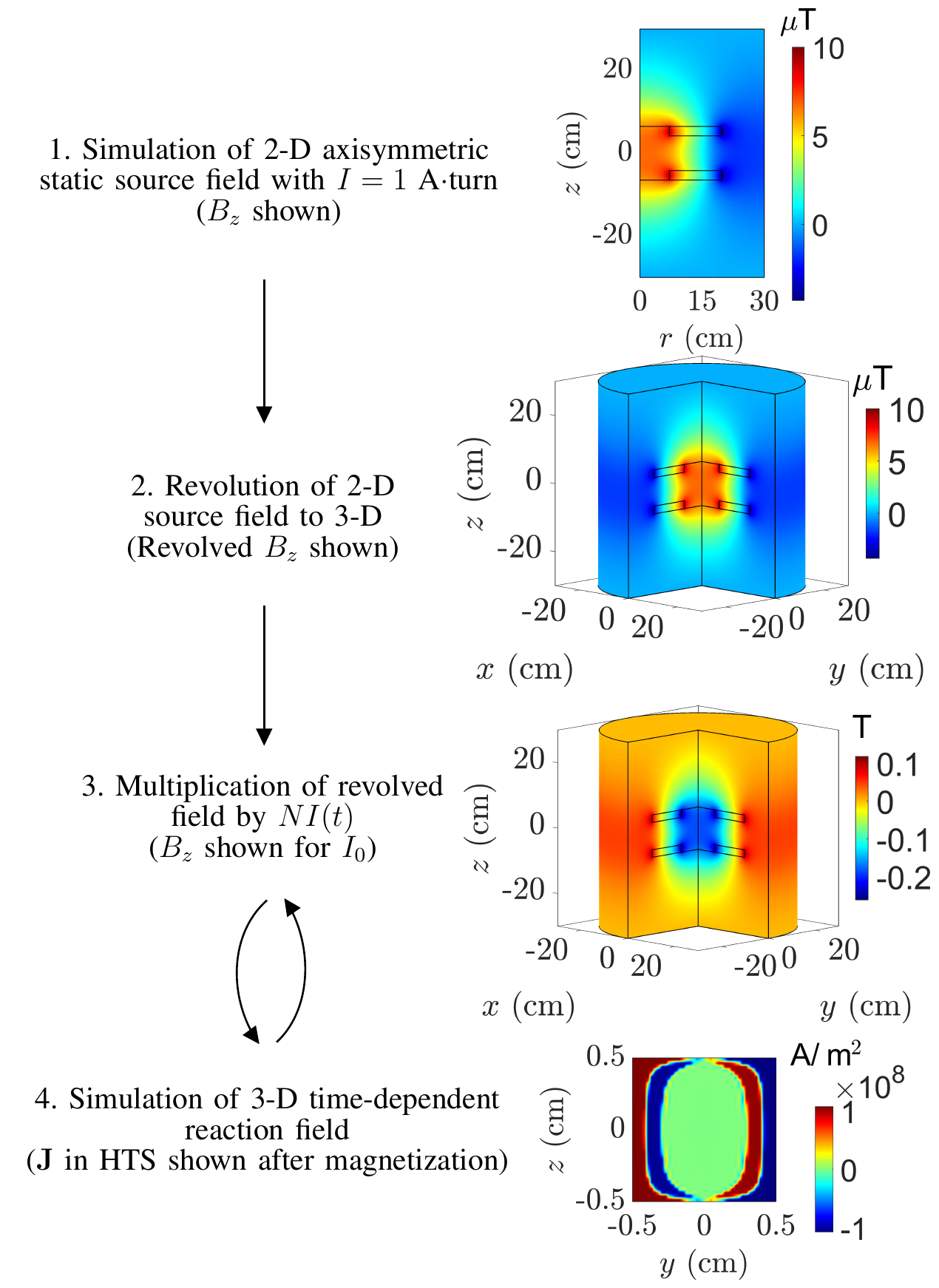}
		\caption{Workflow used to simulate the magnetization of a HTS cube placed between a pair of Helmholtz coils with the \hrphi formulation. Note that the scale of the magnetic flux density in steps 1. and 2. is in the $\mu$T range since a small unitary current of 1~A$\cdot$turn is supplied to the coils.}
		\label{fig:workflow}
	\end{figure}

	The workflow used to simulate the full \threeD magnetization process is shown in Fig.~\ref{fig:workflow}. Four main steps are required: 1) simulating the \twoD axisymmetric, static source field of the coils with a unitary current of $I=1$~A$\cdot$turn, 2) revolving the \twoD normalized source field to \threeD, 3) adjusting the amplitude of the normalized source field according to the time-dependent current inside the coils, and 4) simulating the reaction field of the superconductor due to the source field. Step 1) takes only 6~s and step 2) is done in the preprocessing of steps 3) and 4). Steps 3) and 4) are automatically iterated at each time step in the time-dependent simulation, which takes a total of 25~hours to compute.
	
\begin{figure*}
	\centering
	\includegraphics[width=\linewidth]{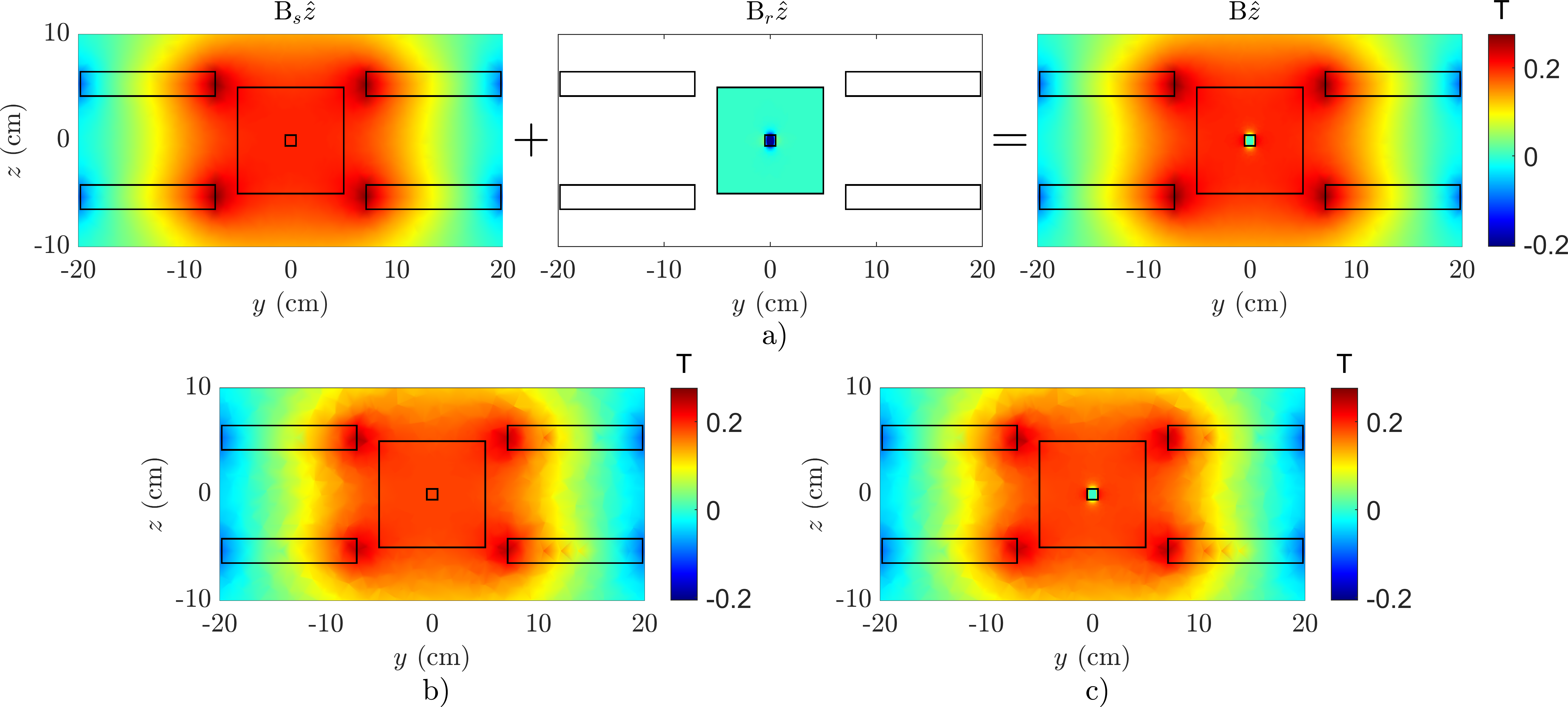}
	\caption{a) Addition of the magnetic flux density in the $z$-direction of the coils ($\mathbf{B}_s$) and of the HTS cube ($\mathbf{B}_r$) in the $y$-$z$ plane calculated with the \hrphi formulation, leading to the total magnetic flux density ($\mathbf{B}$). All figures show the magnetic flux density when the current in the coils first reaches its maximum value. b) Magnetic flux density in the $z$-direction due to the coils calculated with the 3-D H-formulation without the HTS cube. c) Magnetic flux density of the coils and the HTS cube in the $z$-direction calculated with the \threeD H-formulation}
	\label{fig:field}
\end{figure*}

	The principle of the field separation method is illustrated in Fig.~\ref{fig:field}a). In this case, the workflow of Fig.~\ref{fig:workflow} is used along with the \hrphi formulation in order to efficiently model the magnetic response of the HTS cube between the Helmholtz coils. The simulation space of the HTS' reaction field is reduced to a cube of 10~cm sides surrounding the bulk, as shown by the dashed lines in Fig.~\ref{fig:geom}a), since the reaction field has a limited reach. Therefore, the \threeD geometry is identical to that considered in the benchmark, but in this case, the field of the coils is used for the magnetization. 	
	
	Fig.~\ref{fig:field}a) shows that adding the source magnetic flux density (leftmost plot) and reaction magnetic flux density (center plot) yields the total magnetic flux density (rightmost plot) of the system when the current in the coils first reaches $I_0$. The reaction field is only modeled near the HTS cube in order to reduce the number of DOFs, which is why the field is absent (white) far from the bulk. When adding the reaction field to the source field, we find that the diamagnetic nature of the SC is depicted correctly.

	To validate our model, we compare it to the H-formulation simulated in the whole geometry, including the Helmholtz coils and the HTS domain. Several challenges arise when modeling the uniform current density inside the coils in the H-formulation. Firstly, the resistivity of the coils must be set high enough such that eddy currents are not induced, since these would alter the uniform current density needed inside the stranded coils. Additionally, the resistivity must be proportional to 1/$r$ in order for the current to be evenly distributed inside the coils. Since the resistivity of air is usually taken as 1~$\Omega$m in order to avoid eddy currents, and since the radial position of the coils is around 0.1~m, we define the resistivity of the coils as $\rho=0.1/r$, such that it varies from 0.5~$\Omega$m to 1.4~$\Omega$m.

	We found that the usual method of constraining the current in the coils\cite{grilli2013} leads to very long computation times due to the high resistivity of the coils. This can be alleviated by modifying the definition of the current density as \hbox{$\mathbf{J}=\nabla\times\mathbf{H}-\mathbf{J}_{\text{ext}}$}, where $\mathbf{J}_{\text{ext}}$ is the externally applied current density. This method produces a uniform current through the coils, while greatly reducing the computation times in comparison with the constraint method. Indeed, in a \twoD test model, the computation times for our proposed method and the constraint method is 1~minute and 60~minutes, respectively. Note, however, that adding an additional term to the current density to impose a transport current is not suitable when the resistivity in the conductor is low and the conductor is solid instead of stranded, since it does not properly model eddy currents and the skin effect.
	
	For the full \threeD H-formulation simulations, we use a total of 1,027,936 elements in air domains, 33,806 elements in coil domains and 68,921 elements in the HTS domain. With linear curl elements, this corresponds to a total of 1,462,778 DOFs. In comparison, in the \hrphi formulation, we use 374,266 elements in the reduced air domains and 68,921 elements in the HTS domain for the \threeD reaction field simulations. This corresponds to 286,391 DOFs with linear elements, demonstrating an 80\% reduction in the number of DOFs when compared to the full H-formulation. This reduction comes from the use of a scalar dependent variable in air domains and a reduction in the number of mesh elements as a result of the reduced geometry.

	The magnetic flux density in the $y$-$z$ plane simulated with the full \threeD H-formulation is shown in \hbox{Figs.~\ref{fig:field}b) and c)}. In Fig.~\ref{fig:field}b), the HTS cube is omitted to demonstrate the source field obtained when the current first reaches its maximum amplitude of 27.45~kA$\cdot$turn. Fig.~\ref{fig:field}c) shows the complete electromagnetic simulation of the coils and the bulk, demonstrating the expected diamagnetic nature of the HTS cube.
	
	The source magnetic flux density ($\mathbf{B}_s$) and total magnetic flux density ($\mathbf{B}$) of Fig.~\ref{fig:field}a) are very similar to the magnetic flux densities calculated in the full \threeD H-formulation of Fig.~\ref{fig:field}b) and c), respectively. The field is smoother when using the \hrphi formulation because we use quartic elements to compute the \twoD source field nearly instantly, whereas we use linear elements in the full \threeD H-formulation. Fig.~\ref{fig:field}a) demonstrates that adding the reaction field to the source field yields essentially the same result as simulating the full field, but with much faster computation times. Indeed, the full field calculated with the \threeD H-formulation requires 93~hours for a complete magnetization cycle. On the other hand, the full field obtained with the \hrphi formulation requires 25~hours for one magnetization cycle, showing that this method is nearly four times faster than the complete H-formulation.
	
	\begin{figure*}
		\centering
		\includegraphics[width=\linewidth]{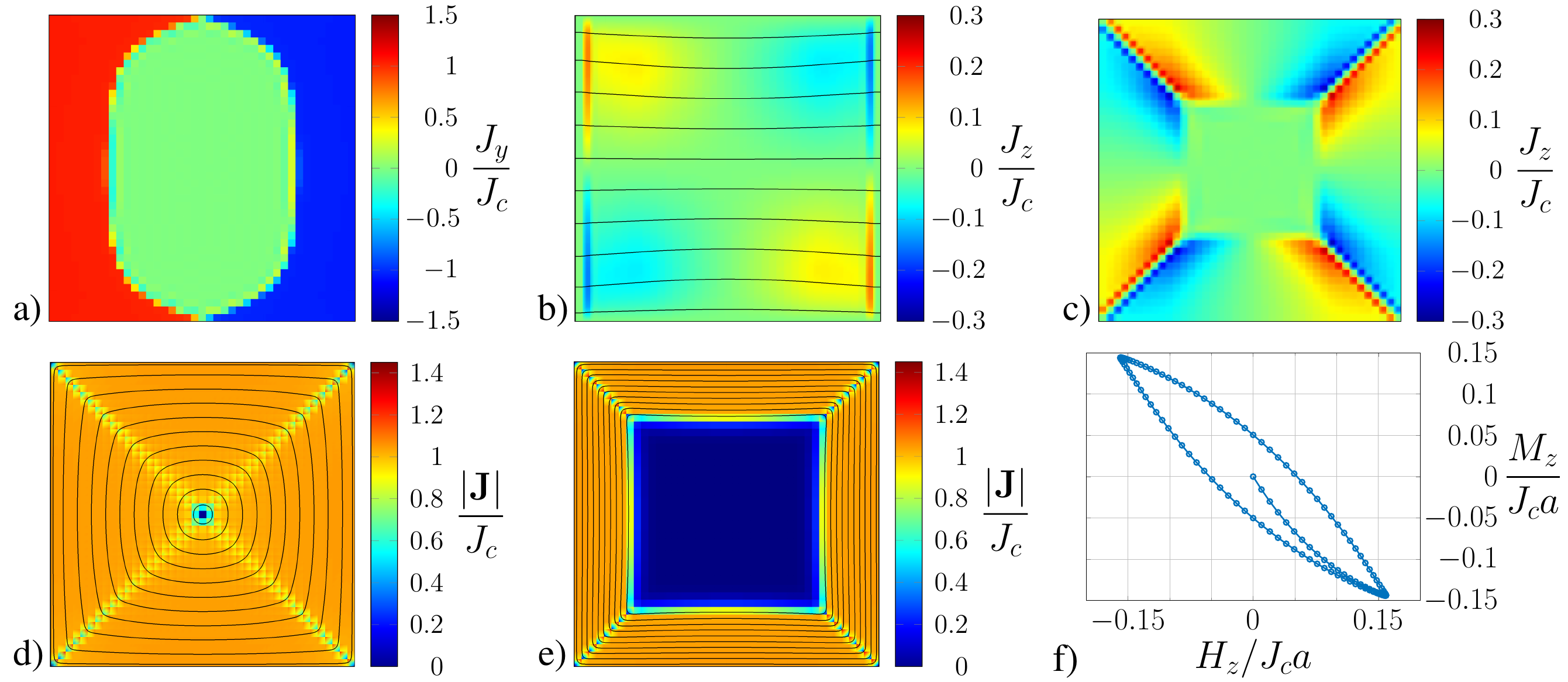}
		\caption{Current density normalized by $J_c$ and magnetization curve of the HTS cube computed with the reduced \hbox{H-$\phi$} formulation. The nearly uniform applied field is generated by the Helmholtz coils. The origin of the coordinate system is taken at the center of the HTS cube. a) Normalized $J_y$ in the $y=0$~mm plane. b) Normalized $J_z$ in the $y=4.88$~mm plane (0.12~mm from the surface of the cube). c) Normalized $J_z$ in the $z=3.9$~mm plane (1.1~mm from the surface of the cube). d) Normalized norm of $\mathbf{J}$ in the $z=4.88$~mm plane (0.12~mm from the surface of the cube). e) Normalized norm of $\mathbf{J}$ in the $z=0$~mm plane. f) Magnetization curve for 1.25 cycle.}
		\label{fig:benchmark}
	\end{figure*}

	Fig.~\ref{fig:benchmark} shows the results of the benchmark problem computed with the reduced \hbox{H-$\phi$} formulation when the current in the magnetizing coils first reaches its maximum amplitude. We find a nearly perfect match with the benchmark for all current density orientations considered in Fig.~\ref{fig:benchmark}a)-e) and for the magnetization curve of Fig.~\ref{fig:benchmark}f). Indeed, when calculating the relative error of the current density between formulations using:
	\begin{equation}
	\epsilon=\left\lvert\frac{\overline{\|J_{H_r-\phi}\|}-\overline{\|J_{H}\|}}{\overline{\|J_{H}\|}}\right\rvert\times 100\%,
	\end{equation}
	we obtain a difference of 1.09\%, where $\overline{\|J_{H_r-\phi}\|}$ and $\overline{\|J_{H}\|}$ are the average values of the norm of $J$ over the superconducting domain calculated with the \hrphi and the H-formulation, respectively. See the htsmodelling.com website for the current densities computed with the H-formulation, the minimum electromagnetic entropy production method \cite{pardo2017} and the volume integral method.

\subsection{Magnetic levitation}

Another application that can greatly benefit from using the reduced \hbox{H-$\phi$} formulation is the magnetic levitation of a permanent magnet above a superconducting bulk, as seen in MagLev systems for example \cite{hyung-woolee2006}. Several methods have been proposed to model this phenomenon \cite{navau2013,sass2015,grilli2018,queval2018, fernandes2020,irina2019}. For example, Sass et al. used a combination of the FEM with an integral method to apply appropriate boundary conditions\cite{sass2015}. They use the Biot-Savart law to calculate the field produced by the permanent magnet and apply this field at the air boundaries very close to the HTS bulk. This yields good results when comparing with experiments, but the integrals required to compute the contribution of the HTS bulk to the boundary conditions are computationally expensive.

Another method proposed by Grilli et al. is to simulate the complete experiment, consisting of a permanent magnet levitating above a HTS bulk by using a moving mesh with the H-formulation \cite{grilli2018}. However, the moving mesh feature is resource intensive and requires careful time-stepping in order to obtain convergence. In addition, a constraint needs to be applied to explicitly impose the divergence-free condition of the magnetic flux density since the moving mesh violates the implicit divergence-free condition of the time-dependent H-formulation, as explained in the Appendix. This issue can be solved by using the \hbox{H-$\phi$} formulation, with the $\phi$ physics in moving domains, such that the divergence-free condition is explicitly defined, but the moving mesh is still computationally expensive.

\begin{table}
	\centering
	\caption{Parameters used for the simulation of the levitation of a PM over a HTS bulk.}
	\begin{tabular}{ l  l  l}
		\toprule
		Parameter & Description & Value \\
		\midrule
		
		$E_{\text{c}}$ & Critical electric field & $1\times10^{-4}~\text{Vm}^{-1}$\\
		$n$ & Power law exponent & 40\\
		$J_{\text{c}}$(0~T) & $J_{\text{c}}$ at 75~K, 0~T & $1.89\times10^8~\text{Am}^{-2}$\\
		$J_{\text{c}}$(0.6~T) & $J_{\text{c}}$ at 75~K, 0.6~T & $1.35\times10^8~\text{Am}^{-2}$\\	
		$\rho_{\text{n}}$ & Normal state resistivity & $1\times10^{-6}~\Omega\text{m}$\\
		$M_0$ & Magnetization of PM & $6.6903\times10^5~\text{Am}^{-1}$\\
		$r_{\text{SC}}$ & Radius of HTS & 12.5~mm\\
		$r_{\text{PM}}$ & Radius of PM & 12.5~mm\\
		$h_{SC}$ & Height of HTS & 18~mm\\
		$g_{\text{ZFC}}$ & Initial gap & 46.81~mm\\
		$d$ & Excursion & 46.71~mm\\
		$v$ & Speed of displacement & 0.38~mm/s\\
		\bottomrule
	\end{tabular}
	\label{tbl:params}
\end{table}

\begin{figure}[t]
	\centering
	\includegraphics[width=\linewidth]{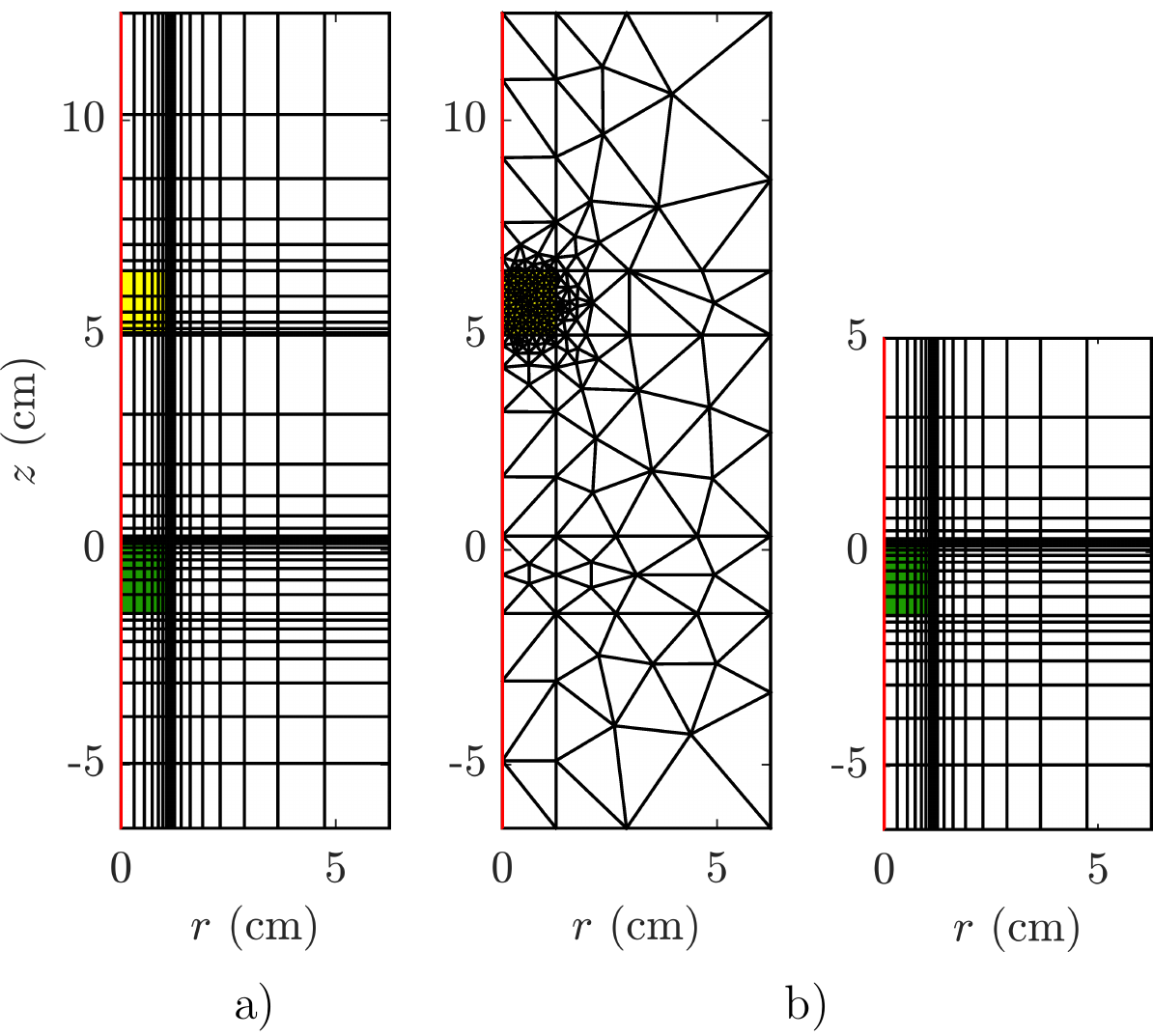}
	\caption{Simulated geometries and meshes considered in the levitation portion of this work. The white, yellow and green domains represent the air, PM and HTS bulk, respectively. a) \twoD axisymmetric geometry considered in the dynamic H-formulation. b) Separated geometry used to simulate the source and reaction fields. The mesh in a) is moved, while both meshes in b) are held fixed. The red line indicates the symmetry axis.}
	\label{fig:geom_levitation}
\end{figure}

In this section, we use the \hrphi formulation to calculate the force generated by a permanent magnet moving towards a HTS bulk without simulating any movement. We consider the same \twoD axisymmetric geometry as in \cite{grilli2018}, shown in Fig.~\ref{fig:geom_levitation}a), where a permanent magnet is initially suspended 46.81~mm above a ZFC HTS bulk. The permanent magnet is lowered over a distance $d=46.71$~mm and brought back up at a constant velocity of $v=0.38$~mm/s. We use a modified power law limited by the normal state resistivity $\rho_\text{n}$ of the HTS to realistically model the over-critical current regime, such that the resistivity is given by:
\begin{equation}
\rho=\frac{\rho_{\text{PL}}\rho_{\text{n}}}{\rho_{\text{PL}}+\rho_{\text{n}}},
\end{equation}
where $\rho_{\text{PL}}$ is given in \eqref{eq:rho} with a field-dependent critical current density at 75~K linearly interpolated from $J_c(0~T)$ to $J_c(0.6~T)$. The relevant parameters used in the simulations are given in Table.~\ref{tbl:params}. The levitation force is calculated from the azimuthal current density $J_{\phi}$ and the radial magnetic flux density $B_{\text{r}}$ induced in the superconductor as:
\begin{equation}
F=\int_S 2\pi rJ_{\phi}B_\text{r}\, dS,
\end{equation}
where $S$ is the cross-section of the HTS in the $r$-$z$ plane.

To simulate the \hrphi formulation, we separate the model into two parts: a time-dependent simulation of the static source magnetic field produced by the permanent magnet and a time-dependent simulation of the reaction field of the HTS generated by the moving source field. Although there are no time-dependent quantities in the source field simulations, they are carried out with a time-dependent solver to introduce the time-dependence of the moving $z$-coordinate of the PM, as explained below. The source field is computed in the whole geometry, while the reaction field is limited to a region surrounding the HTS, as shown in Fig.~\ref{fig:geom_levitation}b). The full geometry is 6.25 x 18.891~cm, while the reduced geometry is 6.25 x 11.481~cm. Note that the mesh is refined around the PM for a more accurate source field calculation, while the mesh used in the reaction field calculation replicates the one used in the dynamic H-formulation of Fig.~\ref{fig:geom_levitation}a) for a fair comparison.

We incorporate the motion of the PM by inputting a time-dependent $z$-coordinate in the extrusion of the source field from the source to the reaction field simulations. In principle, a purely static (stationary in COMSOL terminology) simulation could be carried out for the source field, but introducing the motion of the PM would be less practical in COMSOL because there would be no time parameterization. Nevertheless, the computation time difference between a static and time-dependent source field simulation is only a few seconds.

In order to simulate the reaction field in superconducting domains, we need to employ \eqref{eq:H_sep}. However, in the source field simulations, the magnetic field does not change in time and can therefore not be used in \eqref{eq:H_sep}. To solve this issue, we use the chain rule to introduce the spatial derivative of the magnetic field, so that the governing equation becomes: 
\begin{equation}
\quad \nabla\times \left(\rho\nabla\times\mathbf{H}_r\right) =-\mu_0\left(\frac{\text{d} \mathbf{H}_r}{\text{d} t}+\frac{\text{d} \mathbf{H}_s}{\text{d} z}\frac{\text{d} z}{\text{d}t}\right) \,,
\end{equation}
where d$z$/d$t$ is the displacement velocity of the PM. For the H-formulation and the reaction field simulations, we use quadratic shape functions with 100 mesh elements in the HTS domain. 

To evaluate the performance of the \hrphi formulation with respect to other mixed formulations, we consider the H-A formulation in this \twoD axisymmetric case. Since the magnetic vector potential A is a scalar in \twoD, the number of DOFs in the H-A and H-$\phi$ formulations is equivalent. Therefore, we simulate the reaction field with the H-A formulation (called \hra below) to compare the computation times and accuracies between the two mixed formulations.

	\begin{figure}[t]
	\centering
	\includegraphics[width=\linewidth]{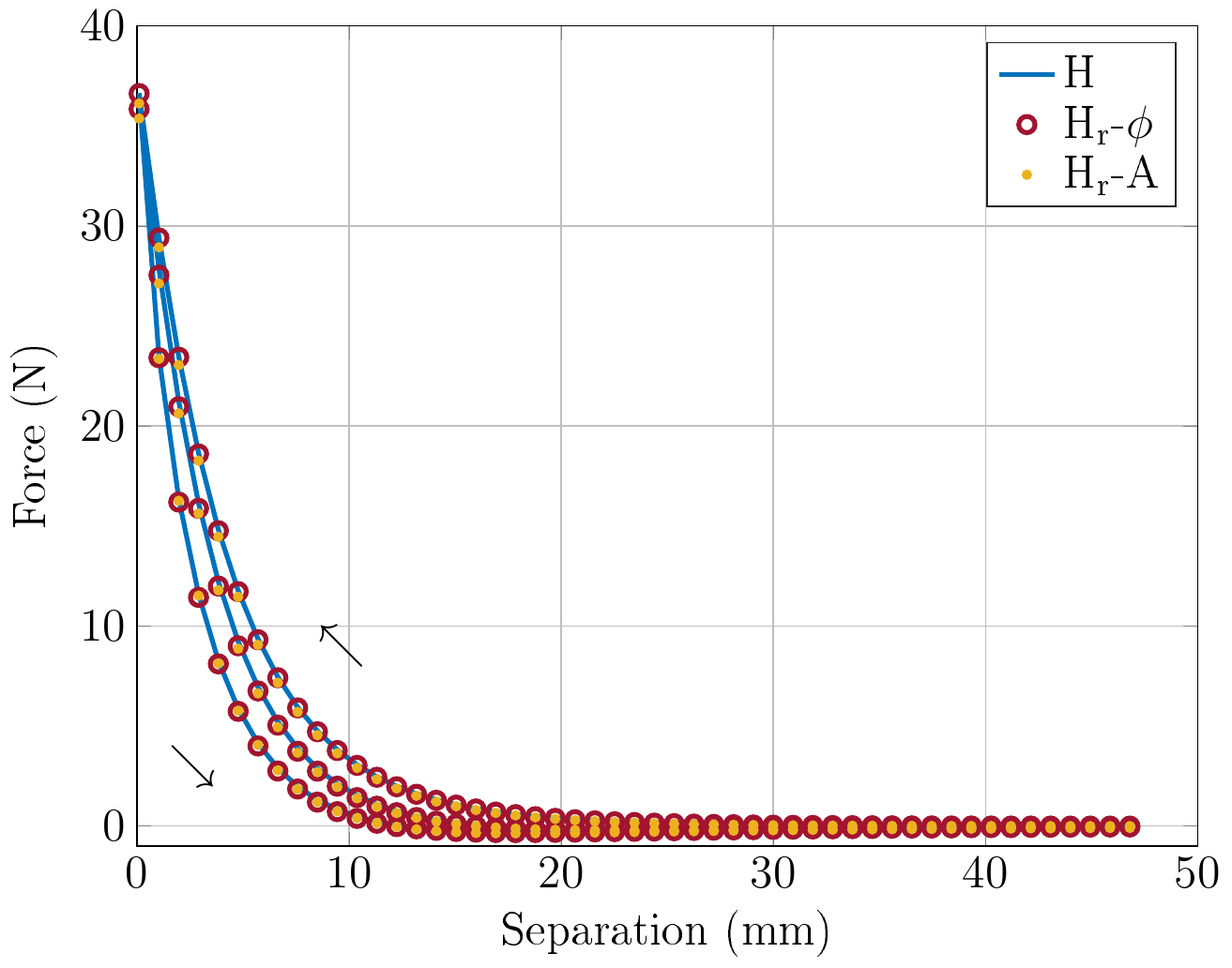}
	\caption{Levitation force versus separation distance between the HTS bulk and the PM obtained with the dynamic H-formulation, the \hrphi formulation, and the \hra formulation. The arrows indicate the path followed by the PM during the first cycle.}
	\label{fig:f_curve}
\end{figure}

The levitation force generated between the HTS bulk and the PM is shown in Fig.~\ref{fig:f_curve} for 1.5 cycle. We find very good agreement between the H, \hrphi, and \hra formulations. However, when comparing the \hrphi and \hra formulations with the H formulation, the percent difference of the force at the shortest separation distance of 0.1~mm is 0.05\% and 1.41\% for the \hrphi and \hra formulations, respectively. Fig.~\ref{fig:f_curve} shows that the \hra formulation slightly underestimates the force when the separation distance is less than 10~mm. The time taken to compute 1.5 cycle is 462~seconds with the H-formulation, while it is only 149~seconds and 165~seconds with the \hrphi and \hra formulations, respectively. The mixed formulations therefore result in similar computation times, while being three times faster than the dynamic H-formulation.

	\begin{figure*}
	\centering
	\includegraphics[width=\linewidth]{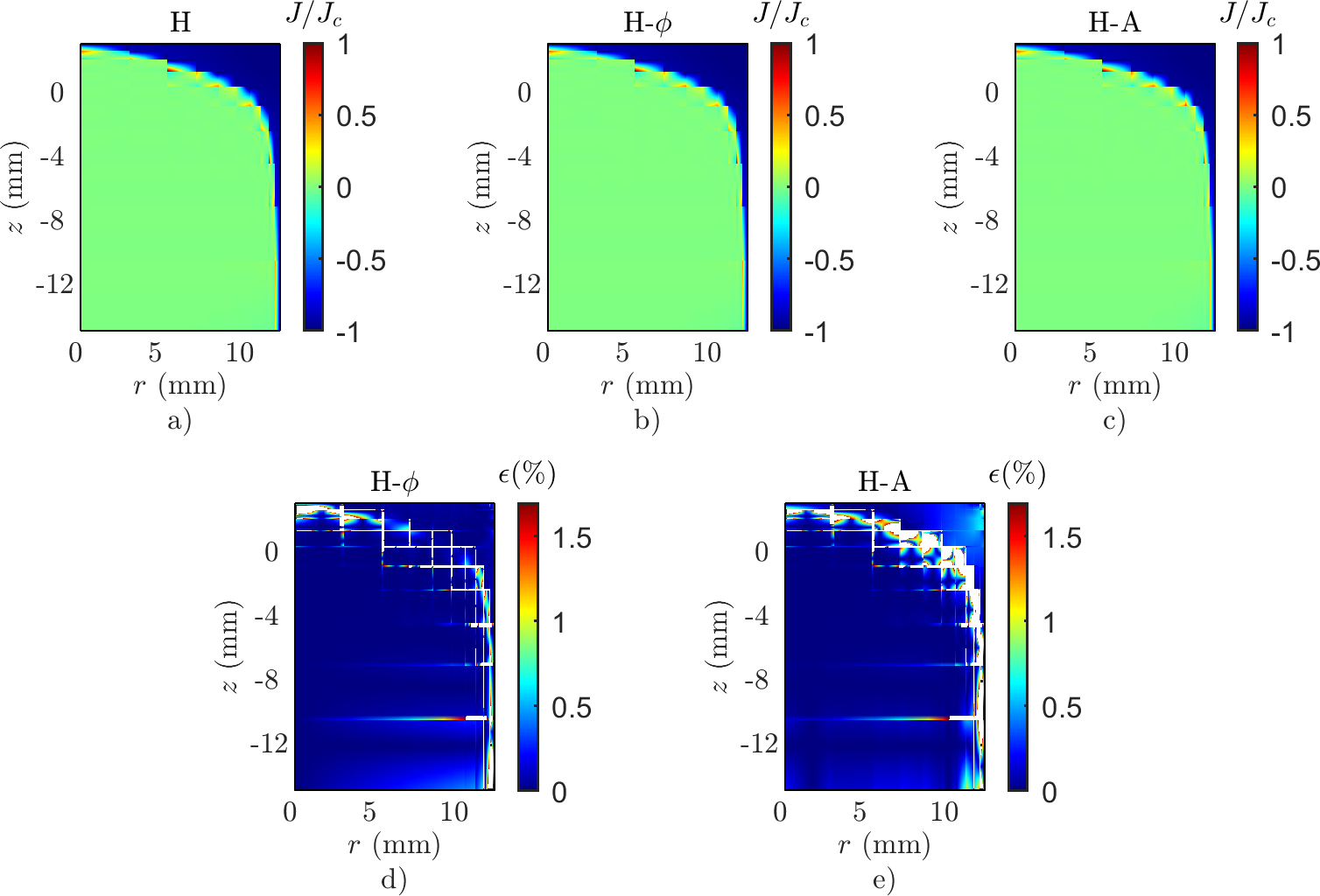}
	\caption{Current density distribution in the HTS bulk when the separation between the HTS and the PM first reaches 0.1~mm. Current density obtained with the H-formulation (a), \hrphi formulation (b) and \hra formulation (c). Percent difference of the current density obtained between the H and \hrphi formulations (d) and between the H and \hra formulation (e). The percent difference $\epsilon$ is only plotted below 1.7\% (higher values are in white) in order to better visualize the areas of interest.}
	\label{fig:J}
\end{figure*}

There are two main reasons why the mixed formulations are faster. First, the number of DOFs are reduced from 10,731 to 2,273. This reduction comes from the use of a scalar dependent variable in air domains, a fixed mesh, a reduced geometry and no Lagrange multipliers to constrain the divergence of the magnetic field to zero. In addition, the absence of a moving mesh enables more lenient time-stepping. Note, however, that the added efficiency of the \hrphi and the \hra formulations comes at the cost of being slightly more complex to implement and post-process than the H-formulation.

A comparison of the current densities obtained with the H, \hrphi and \hra formulations in the HTS bulk when the separation distance between the HTS and the PM first reaches 0.1~mm is shown in Fig.~\ref{fig:J}a)-c). In Fig.~\ref{fig:J}d) and e), we also show the percent difference between the H and \hrphi formulations and the H and \hra formulations, given by:
\begin{equation}
\epsilon=\frac{\left|J_{H_{r}-x}-{J}_{H}\right|}{J_c}\times100\%,
\label{eq:error}
\end{equation}
where $J_{H_{r}-x}$ corresponds to the current density obtained with the \hrphi or the \hra formulation and $J_{H}$ is the current density obtained using the H-formulation. The percent difference is normalized by $J_c$ instead of $J_H$ in order to avoid singularities. We plot $\epsilon$ below 1.7\% to better visualize the distribution in regions of interest.

The most prominent difference in the current density between the formulations is located at the current front. In this region, the current density varies so drastically that a small difference in the position of the front yields a percent difference of up to 147\% for the \hrphi formulation and 151\% for the \hra formulation. However, in areas with significant current density ($>1\times10^8$~A/m$^2$) outside of the current front, the percent difference remains below 1.7\% in all formulations. Comparing Fig.~\ref{fig:J}d) and e), we see that the \hrphi formulation more accurately reproduces the H-formulation results than the \hra formulation, most likely because the coupling between the two physics is more natural in this formulation.

\section{Conclusion}

	In this paper, we described an efficient method of modeling superconductors surrounded by independent magnetic components based on the reduced \hbox{H-$\phi$} formulation. We showed how separating source and reaction fields can reduce computation times by nearly a factor of three and four in \twoD and \threeD, respectively.
	
	We analyzed two models: the magnetization of a HTS cube between a pair of Helmholtz coils and the levitation of a PM over a HTS bulk. In the former, we validated our results with benchmark~\#5 of the htsmodelling.com website\cite{htsmodeling} and showed that the percent difference of the current density between the two formulations is 1.09\%. We also compared computation times with the H-formulation, demonstrating that the \hrphi formulation is nearly four times faster. 
	
	For the levitation simulations, we based our model on previous work by Grilli et al. carried out using the H-formulation and a moving mesh\cite{grilli2018}. By comparing with the H-formulation, we found that the \hrphi and \hra formulations reproduced the levitation force generated between the HTS and the PM with good agreement. We also found that the percent difference of the current density between the formulations remains below 1.7\% in regions of interest. In addition, the \hrphi and \hra formulations were found to be three times faster than the dynamic H-formulation. Finally, when comparing the current densities obtained between formulations, we deduced that the \hrphi formulation more accurately reproduced the H-formulation results than the \hra formulation. Since both mixed formulations require similar computation times, we conclude that the \hrphi formulation is favorable for this particular model.
	
	Although the implementation and post-processing of the reduced \hbox{H-$\phi$} formulation is more challenging than the H-formulation, the computation times are greatly improved. Therefore, this method is well suited to efficiently model large geometries containing many magnetic components surrounding HTS domains.
	
	\section{Acknowledgements} 
	
	This work was supported by the Fonds de recherche du Qu\'ebec --- Nature et Technologies (FRQNT) and TransMedTech Institute and its main funding partner, the Canada First Research Excellence Fund. 
	
	\appendix{\begin{center}Divergence-free condition in moving meshes\end{center}}

	In the following, we show how a moving mesh in the H-formulation causes the implicit divergence-free condition to be lost. When simulating the H-formulation with a fixed mesh, the divergence-free condition of the magnetic flux density is guaranteed at each time step if it is satisfied at the initial time, as shown in \cite{zermeno2013}. This can be seen by taking the divergence of \eqref{eq:H}, so that we obtain:
	\begin{align}
	\nabla\cdot(\nabla\times \left(\rho\nabla\times\mathbf{H}\right)) &=-\nabla\cdot\left(\mu_0\frac{\partial \mathbf{H}}{\partial \text{t}}\right)\\
	\iff \frac{\partial}{\partial \text{t}}\left(\nabla\cdot\mathbf{H}\right)&=0.
	\end{align}
	Thus, if the field is divergence-free initially, it will remain so at every time step.
	
	In the case of a conductor moving in the $z$-coordinate, the magnetic field is given by $H(x,y,z(t),t)$, where the $z$-coordinate depends on time. The numerical derivative of $H$ with respect to $t$ then considers the time-dependence of the $z$-coordinate, meaning that the total derivative is computed. However, in Faraday's law, only the partial derivative of $H$ with respect to time is required. The total derivative is related to the partial derivative through the chain rule:
	\begin{equation}
	\frac{\text{d}H}{\text{d}t}=\frac{\partial H}{\partial t}+\frac{\partial H}{\partial z}\frac{\partial z}{\partial t}.
	\end{equation}

Therefore, in order to numerically calculate the partial derivative with respect to time in \eqref{eq:H}, the H-formulation in a domain moving in the $z$-direction must read:
	\begin{equation}
	\quad \nabla\times \left(\rho\nabla\times\mathbf{H}\right) =-\mu_0\left(\frac{\text{d}\mathbf{H}}{\text{d} t}-\frac{\text{d}\mathbf{H}}{\text{d}z}\frac{\text{d}z}{\text{d}t}\right) \,,
	\end{equation}
	where $\frac{\partial H}{\partial z}\frac{\partial z}{\partial t}=\frac{\text{d} H}{\text{d} z}\frac{\text{d} z}{\text{d} t}$ in this case.	Hence, when taking the divergence of the above equation, we get:
	\begin{equation}
	\nabla\cdot\left(\frac{\text{d}\mathbf{H}}{\text{d} t}-\frac{\text{d}\mathbf{H}}{\text{d}z}\frac{\text{d}z}{\text{d}t}\right)=0 \,,
	\end{equation}
	meaning that we no longer have the condition that $\frac{\partial}{\partial t}\left(\nabla\cdot\mathbf{H}\right)=0$, so the implicit divergence-free condition is lost.
	

	
	
	%
\begingroup
\raggedright
\bibliography{Bibliography}
\bibliographystyle{IEEEtran}
\endgroup
	
	%
	
	\begin{IEEEbiographynophoto}{Alexandre Arsenault}
		received a B.Sc. in physics from McGill University, Montr\'eal, QC, Canada, in 2016. He also received a M.Sc. in physics from McMaster University, Hamilton, ON, Canada, in 2018. He is currently pursuing a Ph.D. degree in biomedical engineering at Polytechnique Montr\'eal under the supervision of Prof. Fr\'ed\'eric Sirois. His research interests include the characterization and simulation of bulk high-temperature superconductors.
	\end{IEEEbiographynophoto}

\begin{IEEEbiographynophoto}{Fr\'ed\'eric Sirois}(S'96--M'05--SM'07)
	received the B.Eng. degree in electrical engineering from Universit\'e de Sherbrooke, Sherbrooke, QC, Canada, in 1997, and the Ph.D. degree in electrical engineering from Polytechnique Montr\'eal, Montr\'eal, QC, Canada, in 2003.
	From 1998 to 2002, he was affiliated as a Ph.D. scholar with the Hydro-Qu\'ebec's Research Institute (IREQ), where he was a Research Engineer from 2003 to 2005. In 2005, he joined Polytechnique Montr\'eal, where he is currently Full Professor. His main research interests are i) the characterization and modeling of electric and magnetic properties of materials, ii) modeling and design of electromagnetic and superconducting devices, and iii) integration studies of superconducting equipment in power systems. He is a regular reviewer for several international journals and conferences.
\end{IEEEbiographynophoto}

	\begin{IEEEbiographynophoto}{Francesco Grilli}
	
	received the M.S. degree in Physics from the University of Genoa, Italy, in 1998, the Ph.D. degree in Technical Sciences from the \'Ecole Polytechnique F\'ed\'erale de Lausanne, Switzerland, in 2004, and the Habilitation in superconductivity for energy applications from the Karlsruhe Institute of Technology, Germany, in 2017.
	
	From 2004 to 2007, he was a Postdoctoral Researcher with the Los Alamos National Laboratory, NM, USA, and from 2007 to 2009, with Polytechnique Montr\'eal, QC, Canada. Since 2009, he has been with the Karlsruhe Institute of Technology, Germany, where he is currently the leader of the group ``AC Losses in High-Temperature Superconductors.'' His main research interests include the 2D and 3D modeling of high-temperature superconductors and the characterization of their properties.
	
	Dr. Grilli was the recipient of the 2008 and 2014 Van Duzer Prize for best contributed non-conference paper published in the IEEE TRANSACTIONS ON APPLIED SUPERCONDUCTIVITY and of the 2011 Dr. Meyer-Struckmann Science Prize for his work on numerical modeling of superconductors.
	
\end{IEEEbiographynophoto}

	
	

\end{document}